# Magnetization Reversal and Nanoscopic Magnetic Phase Separation in Doped $La_{1-x}Sr_xCoO_3$

Joseph E. Davies,[1] J. Wu,[2] C. Leighton,[2,§] and Kai Liu[1,*]

[1] *Department of Physics, University of California, Davis, CA 95616*
[2] *Department of Chemical Engineering and Materials Science, University of Minnesota, Minneapolis, MN 55455*

(9/7/2005)

## Abstract

The doped perovskite cobaltite $La_{1-x}Sr_xCoO_3$ (LSCO) has been advanced as a model system for studying intrinsic magnetic phase separation. We have employed a first-order reversal curve (FORC) method to probe the amount of irreversible switching in bulk polycrystalline LSCO as a function of Sr doping, field cooling procedure, and temperature. The value of the FORC distribution, $\rho$, is used as a measure of the extent of irreversible switching. For $x < 0.18$, the small values of $\rho$, and its ridge-like distribution along local coercivity ($H_c$) and zero bias ($H_b$), are characteristic of non-interacting single domain particles. This is consistent with the formation of an array of isolated nanoscopic ferromagnetic clusters, as observed in previous work. For $x \geq 0.18$, the much larger values of $\rho$, the tilting of its distribution towards negative bias field, and the emergence of regions with negative $\rho$, are consistent with increased long-range ferromagnetic ordering. The FORC distributions display little dependence on the cooling procedure. With increasing temperature, the fraction of irreversible switching determined from the FORC distribution follows closely the ferromagnetic phase fraction measured by La nuclear magnetic resonance. Our results furthermore demonstrate that the FORC method is a valuable first-pass characterization tool for magnetic phase separation.

**PACS number(s): 75.60.-d, 75.30.Kz, 75.60.Jk, 75.50.Cc.**

## I. Introduction

Magnetoelectronic phase separation is a recurring theme in the physics of complex oxides such as cuprates, manganites and cobaltites, and is thought to play a key role in the understanding of some of their most attractive properties, such as high temperature superconductivity (HTS)[1,2] and colossal magnetoresistance (CMR).[3-6] Essentially, the close competition between various ground states with distinct electronic and magnetic properties leads to the spatial coexistence of multiple phases, even in the absence of chemical inhomogeneity.[3,7] Taking manganites as an example, this magnetoelectronic phase inhomogeneity has been observed in many materials systems, using numerous experimental methods. These techniques include direct spatial probes [e.g. scanning tunneling microcopy (STM) and spectroscopy (STS),[8-11] transmission electron microscopy (TEM),[12,13] and magneto-optical imaging[14]], diffraction and scattering techniques [e.g. neutron diffraction and small-angle neutron scattering (SANS)[12,15-20]], resonance techniques [e.g. nuclear magnetic resonance (NMR)[21,22]], as well as numerous less direct probes such as magnetometry,[3-5] transport,[3-5] and electrical noise.[23] It is important to note that this phase separation is also the subject of intense investigation from the theoretical point of view and that its existence can be reproduced with relatively simple models.[5,7,24]

Doped perovskite cobaltites, which have been the subject of far less investigation than their extensively studied manganite counterparts,[12,13,20,25,26] offer some unique opportunities for fundamental investigations of correlated electron oxides. This stems from two important features of perovskite oxides; (i) they possess an additional degree of freedom associated with the Co ion spin state (which cannot be accessed in manganites and cuprates), and (ii) they exhibit a particularly clear and simple form of magnetic phase separation.[12, 13, 25, 26] It has been recently proven by TEM,[12, 13] SANS,[26] and NMR,[27, 28] that at low doping ($x < 0.18$) the canonical doped perovskite cobaltite $La_{1-x}Sr_xCoO_3$ (LSCO) phase separates into ferromagnetic (FM) metallic clusters embedded in a non-FM insulating matrix. As $x$ increases these clusters become more populous leading to a simple coalescence into a long-range ordered FM network and a coincident percolation transition to a metallic state at $x \geq 0.18$.[25, 29] In contrast to many manganite systems this occurs in the absence of any structural phase transition, and the FM and non-FM phases share the same crystal symmetry (LSCO is rhombohedral at $x < 0.30$).[12, 25, 29] This implies that complicating effects due to elastic or magnetoelastic considerations are unlikely to be as important. Moreover, the experimental evidence clearly indicates that the phase inhomogeneity in LSCO occurs on a nanoscopic scale (of the order of 10-30 Å cluster diameters).[26] It is therefore consistent with simple electrostatic considerations for intrinsic magnetoelectronic phase separation.[5, 7] This is clearly different from many manganite systems where the phase separation can occur on mesoscopic length scales.[5, 7]

Indeed, the LSCO system has recently been employed by some of us to elucidate the physical consequences of the existence of this spontaneous nanocomposite.[26] We have found that at $x < 0.18$, where a dense matrix of FM clusters forms in a non-FM matrix, one can observe a giant magnetoresistance-type effect due to field induced alignment of FM clusters.[26] In addition, we have also noted that this situation is analogous to that obtained in relaxor ferroelectrics and that this leads to glassy transport phenomena in this material.[30]

As we have already pointed out, many techniques exist to probe magnetoelectronic phase separation. However, they are often limited by factors such as surface sensitivity and difficulties with preparing pristine surfaces (STM and STS), the need to use neutron sources (neutron diffraction and SANS), or the need for specialized equipment and expertise (NMR). It is highly desirable to have a simple and widely available technique for rapid throughput characterization of the phase-separated state, which can then be complemented with direct probes as required. Here we demonstrate the application of the First-Order Reversal Curve (FORC) method[31-34] as a successful probe of magnetic inhomogeneity in LSCO. The FORC method is a versatile yet simple technique that yields very detailed information about the magnetic characteristics of a sample. It is particularly sensitive to irreversible switching processes during magnetization reversal. For example, we have *quantitatively* determined the onsets and endpoints of irreversible magnetization switching in Co/Pt multilayers[33, 35] and exchange-spring magnets,[34] which deviate significantly from the field values expected from the major loops. The FORC method is also powerful in that it captures distributions of magnetic characteristics, such as switching field distribution (SFD),[34] coercivity distribution,[36, 37] etc.

In this paper we show that FORC is capable of a simple measurement of the extent of irreversible switching in LSCO and that this allows us to clearly distinguish the long range FM ordered regime at $x \geq 0.18$, from the formation of isolated FM clusters at $x < 0.18$. This is despite the fact that simple analysis of the hysteresis loop parameters such as saturation magnetization, coercivity, remnance and saturation field show no clear distinctions between the two regimes. The amount of irreversibility in the magnetization reversal process is measured as a function of doping, temperature and cooling field and compared with NMR and SANS measurements on this system.

## II. Experimental Considerations

The bulk polycrystalline single-phase samples of $La_{1-x}Sr_xCoO_3$ ($0.10 \leq x \leq 0.50$) were prepared by standard solid-state reaction techniques and characterized by x-ray diffraction, scanning electron microscopy, electron microprobe analysis, ac and dc magnetometry, magnetotransport, Co and La NMR, and SANS. The results have been described previously.[25-28, 38] A separate publication will detail the results of scanning TEM investigations showing that no chemical inhomogeneities exist down to 1 nm length scales.[39] The samples used for this particular study had Sr-doping concentrations of $x$ = 0.10, 0.15, 0.18, 0.20, 0.30, 0.40 and 0.50.

FORC measurements were performed using a Princeton Measurements vibrating sample magnetometer (VSM) with a liquid helium continuous flow cryostat for low temperature measurements. The VSM is used to measure a large number (~$10^2$) of first-order reversal curves (FORC's) in the following manner. After saturation, the magnetization $M$ is measured starting from a reversal field $H_R$ back to positive saturation, tracing out a FORC. A family of FORC's is measured at different $H_R$, with equal field spacing, thus filling the interior of the major hysteresis loop. The FORC distribution is then defined by a mixed second order derivative:[31-33]

$$\rho(H_R, H) \equiv -\frac{1}{2} \frac{\partial^2 M(H_R, H)}{\partial H_R \partial H} \quad (1).$$

This eliminates the purely reversible components of the magnetization.[40, 41] Thus, any non-zero $\rho$ corresponds to *irreversible* switching processes. Either a 2-dimensional contour plot or a 3-dimensional plot of the distribution $\rho$ versus $H$ and $H_R$ can then be created to probe details of the magnetization reversal. This is known as a FORC diagram. Alternatively, $\rho$ can be seen as a function of local coercivity $H_c$ and bias field $H_b$ after a coordinate transformation: $H_b = (H + H_R)/2$ and $H_c = (H - H_R)/2$.[31, 32] If a system were composed of a set of independent magnetic particles, the FORC diagram would simply map out the distribution of their coercivity $H_c$ and bias field $H_b$. For real systems, the FORC diagram also contains information about the complex interactions that occur among particles, as will be illustrated later. Thus, FORC diagrams provide much more information than the ensemble average measured by typical magnetic major hysteresis loops.

As mentioned earlier, the FORC distribution, $\rho(H_c, H_b)$, contains information about the distributions of magnetic characteristics. The value of $\rho$ itself, being non-zero, indicates the amount of irreversible switching.[42] If the sample has a weak ferromagnetic component, where the hysteresis loop is narrow and slanted, the magnetization reversal is mostly reversible and a very small $\rho$ is expected; If the sample is a single FM phase with a perfectly square hysteresis loop, then $\rho(H_c, H_b)$ will be a single spot in $(H_c, H_b)$ space with a value of unity. Furthermore, the projection of $\rho$ onto the $H_b$ axis, in essence an integration, $\rho(H_b)$, characterizes the distribution in bias or interaction field strength, which in turn is affected by such parameters as the proximity of constituents and interaction mechanisms (exchange, dipolar, etc); the projection of $\rho$ onto the $H_c$ axis, $\rho(H_c)$, characterizes the coercivity distribution, which in turn is determined by such parameters as average constituent size and anisotropy. By integrating $\rho(H_c, H_b)$ over the entire $(H_c, H_b)$ space, we capture the total fraction of the sample that has irreversibly switched, $M_{IRREV}$. An approximation for this can be obtained by summing the value of $\rho$ over the entire dataset and then multiplying by the step sizes in $H_b$ and $H_c$:

$$M_{IRREV} = \int \rho(H_c, H_b) dH_c dH_b \approx \sum \rho(H_c, H_b) \Delta H_c \Delta H_b \qquad (2)$$

This value can be used to compare how the amount of irreversible magnetization varies with sample (different Sr doping) and under different conditions (cooling fields, temperatures, etc.). It is important that the data is properly scaled, or normalized, to ensure meaningful comparisons between measurements. Finally we will compare the amount of irreversible magnetization determined from the FORC measurements to values from previous La NMR data.[28]

### III. Results

#### A. Effect of Sr doping

The FORC's, along with the major hysteresis loops, were first measured for each of the different Sr doping levels in LSCO. These measurements were carried out after zero-field cooling to 35 K and a nominal field step size of 25 Oe was used during the FORC measurement. Figure 1 shows the reversal curves mapping out the interior of the major loop (left panel) with the plot of the corresponding FORC distribution $\rho$ plotted using the $H_c - H_b$ coordinates (right panel). In order to compare the samples, the magnetization $M$ (emu/g) was used to compute the FORC distribution $\rho$ (in arbitrary units), where the integration $\int \rho(H_c, H_b) dH_c dH_b$ recovers $M$. Thus small concentrations of FM phases also correspond to small values of $\rho$. Additionally, the contour plots map out the coercivity and bias field distributions, which are characteristic of each sample. The contours were colored on a scale determined by the maximum value of the FORC distribution for the $x = 0.50$ sample, which has the highest saturation magnetization $M_s$. Using the same contour weightings for all samples allows us to clearly visualize the changes in FM behavior as the doping is varied.

For the lowest doping level, $x = 0.10$, the FORC distribution (Fig. 1b) is relatively featureless in comparison to the other samples. The maximum $\rho$ is $\sim 7.2 \times 10^{-7}$, roughly two orders of magnitude smaller than the other samples. This indicates little irreversible switching in this sample, and thus a rather small amount of FM interactions. This reflects well the small saturation magnetization (0.76 emu/g) and the slanted major loop shown in Fig. 1a (delineated by the outer boundary of the FORC's). These findings are consistent with the fact that this composition is far from the $x = 0.18$ critical composition for long-range FM ordering. The system is therefore dominated by non-ferromagnetic phases, with any residual structure in the FORC being attributed to the small amount of ferromagnetic phase fraction.

As $x$ increases to 0.15, strikingly different patterns are seen (Figs. 1c & 1d). The FORC's show a large coercivity of 1.4 kOe and an appreciable increase in the saturation magnetization to $M_S = 10$ emu/g. (The saturation magnetization is determined using the conventional method of extrapolating the linear slope of the major loop at high fields to zero field and taking the intersection.) The FORC distribution $\rho$ has a clear feature in the form of a peak stretching along $H_c$ and centered about $H_b = 0$. The peak has a maximum value of $1.1 \times 10^{-5}$, located at $H_c = 1.4$ kOe and $H_b = 0$. This type of feature is typical of assemblies of single domain nanoparticles where the spread in $\rho$ along $H_c$ corresponds to the distribution of coercivities in a sample (due to the particle size distribution) and the spread in $\rho$ along $H_b$ indicates the distribution of bias fields (i.e. the amount of inter-particle interaction).[37] The pattern shown in Fig. 1d agrees well with the clustered phase described earlier where short ranged FM clusters exist in a non-ferromagnetic matrix.[27] The large aspect ratio of the peak, and the large spread in $H_c$ vs. small spread in $H_b$, indicate that the FM clusters are largely non-interacting. Furthermore, the average cluster size, previously reported to be on the order of 10 – 30 Å at low temperatures,[26] is related to the peak position in $H_c$. The much larger average coercivity compared to the other samples suggests that the clusters are large enough to withstand thermal fluctuations, but are still isolated so that the coercivity is enhanced due to magnetization reversal by rotation. This is analogous to the maximum coercivity enhancement seen in certain size fine magnetic particles: the coercivity decreases in larger particles due to the formation of a multi-domain state, but decreases in smaller particles due to thermal fluctuations.[43, 44]

When the doping level is increased beyond $x = 0.18$, the saturation magnetization increases gradually to 18.7, 20.3, 21.1, and 25.1 emu/g at 35 K for $x = 0.18$, 0.20, 0.30, and 0.50, respectively. The FORC's (Figs. 1e & 1g) show a reduction in the major loop coercivity compared to $x = 0.15$. This indicates that the clusters have coalesced in to a long-range ordered percolated network, leading to multi-domain type reversal and a reduction in the pinning by the non-magnetic phase fraction. The FORC distribution $\rho$, instead of having a feature centered on $H_b = 0$ Oe, is now tilted towards negative $H_b$ at larger values of $H_c$ (Figs. 1f & 1h). The tilt angle is ~45°, thus the pattern is aligned with the $H_R$-axis if transformed into the $H$-$H_R$ coordinates. The peak maximum for $x=0.30$ has a value of $7.2\times10^{-5}$, located at $H_c = 0.5$ kOe and $H_b = -0.2$ kOe (Fig. 1f), and that for $x=0.50$ has a value of $1.2\times10^{-4}$, located at $H_c = 0.2$ kOe and $H_b = -0.1$ kOe (Fig. 1h). The more pronounced peak in the FORC distribution at higher Sr-doping is consistent with a more dominant FM phase fraction. The negative bias fields are clear signatures of increased interactions amongst different magnetic regions of the sample. They have been seen arising from FM-FM exchange coupling in spring magnets[34] and FM-AF exchange coupling in exchange biased films.[45] More subtly, there are small regions of negative $\rho$ next to the positive peaks in Figs. 1f & 1h (dark blue region, online graph). Such negative/positive pairings of FORC features have often been observed in samples where magnetization reversal involves domain formation and wall motion.[33, 37] In those cases, the $H_R$-dependent susceptibility decreases/increases as the domain state responds differently to the applied field, giving rise to the negative/positive $\rho$ pairing. Hence, the large $M_s$, the outstanding FORC peak, the tilting of the FORC pattern to negative bias, along with the existence of negative/positive $\rho$ pairing, are strong indications that the FM clusters have coalesced and that long-range ferromagnetic ordering now dominates.

Integration of the FORC distribution shown in Fig. 1 using equation (2) allows us to probe the evolution of the fractional amount of irreversible magnetization $M_{IRREV}$ as the doping is increased. As noted earlier, the computed $M_{IRREV}$ is an approximation. The qualitative variation of $M_{IRREV}$ with doping, as shown in Fig. 2, is more meaningful than the absolute values of $M_{IRREV}$, which can be scaled and will be shown later. The left-axis of Fig. 2 shows the amount of irreversible magnetization (in emu/g) obtained from the integration. As expected $M_{IRREV}$ is nearly zero at the lowest doping ($x = 0.10$), rapidly increases at $x = 0.15$, and then starts to level off beyond $x = 0.18$. At $x = 0.50$, $M_{IRREV}$ is about 16.5 emu/g, or 66% of its $M_S$ of 25.1 emu/g. The right-axis of Fig. 2 shows the normalized $M_{IRREV}$ obtained by scaling it to 25.1 emu/g, the $M_S$ of the $x = 0.50$ sample. This plot clearly shows the dominance of the FM phase fraction with increased doping.

**B.     Effect of field cooling**

We have also studied whether the procedure by which we cool to 35 K has any impact on the reversal behavior. Previous work has shown an appreciable difference between the zero-field cooled (ZFC) and field cooled (FC) spontaneous magnetizations.[25] While most of those measurements were made in a cooling field of only 10 Oe, we maximize our chance of observing any difference in the switching behavior by cooling in a large field of 18 kOe. Besides the applied cooling field, these FORC measurements were made using the same parameters as the ZFC measurements.

Initial inspection of the contour plots of the FORC distribution showed little appreciable difference between the ZFC and FC measurements. To make a closer comparison between the two measurements, we project the FORC distribution onto both $H_b$ and $H_c$ axes. The projections onto $H_b$ ($H_c$) are done by taking vertical (horizontal) cuts of the FORC distribution separated by the 25 Oe field spacing (the field step for the FORC measurement), spanning the entire dataset, and then summing them to effectively integrate the distribution over $H_c$ ($H_b$). As mentioned earlier, the projection onto $H_b$ represents the distribution of bias fields and is indicative of the strength of interaction in the sample. Projecting onto $H_c$ represents the distribution of coercivities and therefore the size distribution of the magnetic constituents. Here the FORC distribution $\rho$ is computed using $M/M_S$, instead of just $M$ as in Fig. 1, to better illustrate the features in $\rho$–projections within each sample.

The projections of the FORC distribution onto $H_b$ and $H_c$ are shown in Fig. 3 for the samples presented in Fig. 1. It is clear that the projections for the FC measurements very closely follow those of the ZFC. Therefore the effect of field cooling is negligible in terms of the magnetization reversal behavior of the samples. This is despite the previous observation that small cooling fields have distinct effects on the actual magnetization.[25] The projections $\rho(H_b)$ do clearly show unbiased distributions for $x =$ 0.10 and 0.15 (Figs. 3a & 3c) and negatively biased distributions for $x = 0.30$ and 0.50 (Figs. 3e & 3g). The projections $\rho(H_c)$ show the rapid decrease of coercivity for $x \geq 0.15$ (Figs. 3d, 3f, and 3h). These are consistent with the formation of FM clusters in a non-magnetic matrix for $x < 0.18$ and the coalesence into a long-range ordered FM phase for $x \geq 0.18$.

C. **Effect of Temperature**

We have also studied how increasing the temperature affects the reversal behavior and the corresponding FORC distribution. Previous work with La NMR and SANS has allowed us to measure the non-trivial temperature dependences of the FM phase fraction and the average FM cluster size, information that could be correlated with the FORC results.[26-28] We measured the FORC distributions for one sample in the clustered phase with $x = 0.15$ (Fig. 4) and one sample in the long-range ordered FM phase with $x = 0.30$ (Fig. 6). The measurements were made at temperatures ranging from 35 K and up for each sample. Note that here $M/M_S(T)$ was used to compute the FORC distribution $\rho$ at each temperature, where the integration $\int \rho(H_c, H_b) dH_c dH_b$ recovers $M/M_S(T)$. This allows us to separate out the temperature dependence of $M_S$ when evaluating the change of irreversible switching behavior.

For $x$ = 0.15, the main feature of the FORC distribution is spread over roughly 3 kOe along $H_c$ at 35 K (Fig. 4b), but rapidly becomes narrower with a spread of less than 1 kOe at 60 K (Fig. 4d). Correspondingly, the major loops for these samples show a drastic reduction in the coercivity (Figs. 4a, 4c, & 4e). Projecting the FORC distribution onto $H_b$ (Fig. 5a) shows a monotonic decrease in the peak-width while remaining centered on 0 Oe. This indicates that the already limited cluster-cluster interaction becomes even weaker and eventually vanishes near 150 K. This is in remarkable agreement with SANS data showing that the first indications of FM cluster nucleation occur at 150 K for $x$ = 0.15.[26] The projection onto $H_c$ (Fig. 5b) shows that the peak moves closer to $H_c$ = 0 Oe, again with a monotonic decrease of the peak-width. Hence, the FORC distribution becomes localized about the origin, showing the weakening of FM ordering within the clusters.

For $x$ = 0.30, the sample exhibiting long-range order, the FORC distribution $\rho$ approaches the origin more gradually, maintaining the downward curve in the contour plots until 200 K (Fig. 6). With increasing temperature, the *negative bias* in $\rho(H_b)$ decreases, along with a monotonic decrease of the peak-width (Fig. 7a). This trend is in contrast to that of the *unbiased* $\rho(H_b)$ shown in Fig. 5a, indicating an overall stronger interaction amongst constituents. The coercivity and its spread both decrease at higher temperatures, and finally disappear above 200 K (Fig. 7b), indicating the expected FM – paramagnetic transition.

As the features for both samples become more localized, the amount of irreversible magnetization is expected to decrease with increasing temperatures. This is indeed observed by integrating over the FORC distribution for each sample using equation 2 (Fig. 8). Here the $M_{IRREV}$ is normalized to the saturation magnetization at each temperature $M_S(T)$. The decrease is rather drastic for the $x = 0.15$ sample (Fig. 8a) with the fraction of irreversible magnetization leveling off at a very small value between 100 K and 150 K, again consistent with earlier SANS result.[26] Interestingly, the $x = 0.30$ sample (Fig. 8b) shows a slight increase in $M_{IRREV}$ prior to a more convex decrease to near zero at 210 K. This will be discussed below. Note that previous measurements indicate a bulk $T_C$ of 220 K.[25]

The data for $M_{IRREV}/M_S$ versus temperature are plotted along with previous La NMR data[28] showing the fraction of FM-phase material at different temperatures. Unfortunately $x = 0.15$ NMR data is not available so the FORC distribution is compared against the NMR data for $x = 0.10$ (Fig. 8a), i.e. another sample in the isolated cluster regime below $x = 0.18$. For the $x = 0.10$ composition the $M_{IRREV}/M_S(T)$ values determined from FORC are lower, and reach their minimum value at a lower temperature, than the FM-phase fraction from NMR. In this case it is clear that the extent of irreversible switching is not closely related to the ferromagnetic phase fraction, likely due to the fact that the magnetization reversal involves significant amounts of reversible processes. This is consistent with the above conclusions on the nature of the magnetization reversal in the clustered state. For the $x = 0.30$ sample, it is remarkable that the same general shape between the NMR and FORC data has been observed. This suggests that the amount of irreversible switching correlates very well with the amount of FM phases, meaning that conventional irreversible magnetization reversal mechanisms (i.e. domain processes) dominate when a percolated FM state prevails. The turnover at low temperatures is consistent with the La NMR data and was interpreted in terms of a thermally induced phase conversion from glassy to FM phases due to a spin-state transition near 80 K.[28] The decrease in the amount of irreversible switching at higher temperatures is consistent with the bulk $T_C$ of 220 K. It is noteworthy that $M_{IRREV}/M_S$ falls below the La NMR data at high T and that the NMR data appear to indicate that the FM phase fraction persists above the bulk $T_C$. This is due to the existence of isolated FM clusters even above $T_C$, which do not contribute to the amount of irreversible switching.

## IV. Conclusions

We have studied a series of magnetically phase separated $La_{1-x}Sr_xCoO_3$ samples with different Sr doping utilizing the FORC method. FORC has given a detailed account of the magnetization reversal behavior as well as tracing the amount of irreversible switching. Three different comparisons of the FORC data were presented. First, the FORC distributions were compared for samples with different doping, clearly showing the transition from a clustered state ($x < 0.18$) to long range ferromagnetic ordering ($x > 0.18$). The clustered state is characterized by much smaller values of the FORC distribution $\rho$, which is centered at $H_b = 0$ Oe. Ridge-like features extending along $H_c$ at $H_b = 0$ Oe are characteristic of non-interacting single-domain particles, further supporting the previously proposed "cluster-model". As the Sr-doping is increased and FM ordering becomes long-range the values of $\rho$ become larger, indicating a higher degree of irreversible magnetization. Also, the main feature in the contour plots tilts to negative $H_b$ and $\rho$ becomes negative in some regions. The second measurement showed that the effect of field cooling was negligible when looking at the reversal behavior of the samples. Projections of the FORC distribution onto $H_b$ and $H_c$ clearly show this null effect.

Finally, the temperature dependence of samples in the clustered and long-range FM phases was observed. Integration of the FORC distribution shows an overall decrease in the amount of irreversible magnetization with increasing temperatures, consistent with the change of FM-phase fractions measured by NMR. These results further demonstrate that the FORC method is an effective tool to study the magnetic properties of phase separated magnetic materials such as cobaltites and manganites.

**Acknowledgements**

This work has been supported by NSF (EAR-0216346) and the University of California (CLE). Work at UMN was supported by the donors of the American Chemical Society Petroleum Research Fund and the NSF through DMR-0509666. We thank R. K. Dumas, C. R. Pike, K. L. Verosub, R. T. Scalettar, and G. T. Zimanyi for helpful discussions.

**References**

**References**

§   Electronic address: leighton@tc.umn.edu.

*   Electronic address: kailiu@ucdavis.edu.

[1]  G. Burns, *High Temperature Superconductivity - An Introduction* (Academic, Boston, 1992).

[2]  R. Wesche, *High Temperature Superconductors: Material Properties and Applications* (Kluwer Academic, Boston, 1998).

[3]  Y. Tokura and Y. Tomioka, J. Magn. Magn. Mater. **200**, 1 (1999).

[4]  J. M. D. Coey, M. Viret, and S. von Molnar, Ad. Phys. **48**, 167 (1999).

[5]  E. Dagotto, *Nanoscale phase separation and colossal magnetoresistance* (Springer, New York, 2002).

[6]  K. Liu, X. W. Wu, K. H. Ahn, T. Sulchek, C. L. Chien, and J. Q. Xiao, Phys. Rev. B **54**, 3007 (1996).

[7]  E. Dagotto, T. Hotta, and A. Moreo, Phys. Rep.-Rev. Sec. Phys. Lett. **344**, 1 (2001).

[8]  M. Fath, S. Freisem, A. A. Menovsky, Y. Tomioka, J. Aarts, and J. A. Mydosh, Science **285**, 1540 (1999).

[9]  S. F. Chen, P. I. Lin, J. Y. Juang, T. M. Uen, K. H. Wu, Y. S. Gou, and J. Y. Lin, Appl. Phys. Lett. **82**, 1242 (2003).

[10]  T. Becker, C. Streng, Y. Luo, V. Moshnyaga, B. Damaschke, N. Shannon, and K. Samwer, Phys. Rev. Lett. **89**, 237203 (2002).

[11]  C. H. Renner, G. Aeppli, B.-G. Kim, Y.-A. Soh, and S. W. Cheong, Nature **416**, 518 (2002).

[12] R. Caciuffo, D. Rinaldi, G. Barucca, J. Mira, J. Rivas, M. A. Senaris-Rodriguez, P. G. Radaelli, D. Fiorani, and J. B. Goodenough, Phys. Rev. B **59**, 1068 (1999).

[13] J. Mira, J. Rivas, G. Baio, G. Barucca, R. Caciuffo, D. Rinaldi, D. Fiorani, and M. A. S. Rodriguez, J. Appl. Phys. **89**, 5606 (2001).

[14] M. Tokunaga, Y. Tokunaga, and T. Tamegai, Phys. Rev. Lett. **93**, 037203 (2004).

[15] J. W. Lynn, R. W. Erwin, J. A. Borchers, Q. Huang, A. Santoro, J. L. Peng, and Z. Y. Li, Phys. Rev. Lett. **76**, 4046 (1996).

[16] Q. Huang, J. W. Lynn, R. W. Erwin, A. Santoro, D. C. Dender, V. N. Smolyaninova, K. Ghosh, and R. L. Greene, Phys. Rev. B **61**, 8895 (2000).

[17] M. Hennion, F. Moussa, G. Biotteau, J. Rodriguez-Carvajal, L. Pinsard, and A. Revcolevschi, Phys. Rev. Lett. **81**, 1957 (1998).

[18] G. R. Blake, L. Chapon, P. G. Radaelli, D. N. Argyriou, M. J. Gutmann, and J. F. Mitchell, Phys. Rev. B **66**, 144412 (2002).

[19] C. Simon, S. Mercone, N. Guiblin, C. Martin, A. Brulet, and G. Andre, Phys. Rev. Lett. **89**, 207202 (2002).

[20] R. Caciuffo, J. Mira, J. Rivas, M. A. Senaris-Rodriguez, P. G. Radaelli, F. Carsughi, D. Fiorani, and J. B. Goodenough, Europhys. Lett. **45**, 399 (1999).

[21] G. Papavassiliou, M. Fardis, M. Belesi, T. G. Maris, G. Kallias, M. Pissas, D. Niarchos, C. Dimitropoulos, and J. Dolinsek, Phys. Rev. Lett. **84**, 761 (2000).

[22] M. Bibes, L. Balcells, S. Valencia, J. Fontcuberta, M. Wojcik, E. Jedryka, and S. Nadolski, Phys. Rev. Lett. **87**, 067210 (2001).

[23] B. Raquet, A. Anane, S. Wirth, P. Xiong, and S. von Molnar, Phys. Rev. Lett. **84**, 4485 (2000).

[24] K. H. Ahn, T. Lookman, and A. R. Bishop, Nature **428**, 401 (2004).

[25] J. Wu and C. Leighton, Phys. Rev. B **67**, 174408 (2003).

[26] J. Wu, J. W. Lynn, C. J. Glinka, J. Burley, H. Zheng, J. F. Mitchell, and C. Leighton, Phys. Rev. Lett. **94**, 037201 (2005).

[27] P. L. Kuhns, M. J. R. Hoch, W. G. Moulton, A. P. Reyes, J. Wu, and C. Leighton, Phys. Rev. Lett. **91**, 127202 (2003).

[28] M. J. R. Hoch, P. L. Kuhns, W. G. Moulton, A. P. Reyes, J. Lu, J. Wu, and C. Leighton, Phys. Rev. B **70**, 174443 (2004).

[29] M. A. Senaris-Rodriguez and J. B. Goodenough, J. Solid State Chem. **118**, 323 (1995).

[30] J. Wu, J. C. Burley, H. Zheng, J. F. Mitchell, and C. Leighton, unpublished.

[31] C. R. Pike, A. P. Roberts, and K. L. Verosub, J. Appl. Phys. **85**, 6660 (1999).

[32] H. G. Katzgraber, F. Pazmandi, C. R. Pike, K. Liu, R. T. Scalettar, K. L. Verosub, and G. T. Zimanyi, Phys. Rev. Lett. **89**, 257202 (2002).

[33] J. E. Davies, O. Hellwig, E. E. Fullerton, G. Denbeaux, J. B. Kortright, and Kai Liu, Phys. Rev. B **70**, 224434 (2004).

[34] J. E. Davies, O. Hellwig, E. E. Fullerton, J. S. Jiang, S. D. Bader, G. T. Zimanyi, and Kai Liu, Appl. Phys. Lett. **86**, 262503 (2005).

[35] M. S. Pierce, C. R. Buechler, L. B. Sorensen, J. J. Turner, S. D. Kevan, E. A. Jagla, J. M. Deutsch, T. Mai, O. Narayan, J. E. Davies, K. Liu, J. H. Dunn, K. M. Chesnel, J. B. Kortright, O. Hellwig, and E. E. Fullerton, Phys. Rev. Lett. **94**, 017202 (2005).

[36] C. R. Pike, C. A. Ross, R. T. Scalettar, and G. Zimanyi, Phys. Rev. B **71**, 134407 (2005).

[37] R. K. Dumas, K. Liu, I. V. Roshchin, C. P. Li, and I. K. Schuller, unpublished.

[38] M. J. R. Hoch, P. L. Kuhns, W. G. Moulton, A. P. Reyes, J. Wu, and C. Leighton, Phys. Rev. B **69**, 014425 (2004).

[39] C. Perrey, J. Wu, J. S. Parker, C. B. Carter, and C. Leighton, unpublished.

[40] F. Preisach, Z. Phys **94**, 227 (1935).

[41] I. D. Mayergoyz, *Mathematical Models of Hysteresis* (Springer - Verlag, New York, 1991).

[42] This is true for $H_c>0$. Certain reversible components may be manifested in a FORC distribution along $H_c=0$, as discussed in Ref. 36.

[43] K. Liu and C. L. Chien, IEEE Trans. Magn. **34**, 1021 (1998).

[44] B. D. Cullity, *Introduction to Magnetic Materials (Addison-Wesley Series in Metallurgy and Materials*, 1972).

[45] K. Liu, et al., unpublished.

**Figure captions**

**Fig. 1 (color online):** FORC's (a, c, e, g) along with corresponding contour plots of the FORC distributions (b, d, f, h, respectively) of LSCO samples with $x$=0.10, 0.15, 0.30 and 0.50 obtained at 35 K. The first point of each FORC is represented by a black dot. Contour coloring scale is the same for all contour plots with the maximum value (red) being normalized to the maximum value of the $x = 0.50$ sample. The maximum value of the FORC distribution $\rho$ is 7.2 x $10^{-7}$, 1.1x$10^{-5}$, 7.2 x $10^{-5}$, and 1.2 x $10^{-4}$ for (b), (d), (f), and (h), respectively, as shown in the inset legend of (b).

**Fig. 2: (color online):** The fractional amount of irreversible magnetization $M_{IRREV}$ at 35 K as determined by integration of the FORC distribution $\rho$ shown in Fig. 1 using equation (2). The $y_1$-axis shows $M_{IRREV}$ in emu/g; the $y_2$-axis shows normalized $M_{IRREV}$ as a percentage of the saturation magnetization for the $x$=0.50 sample. Solid line is a guide to the eye.

**Fig. 3:** Projections of the FORC distribution at 35 K onto $H_b$ (right column) and $H_c$ (left column) for both the zero-field cooled (ZFC, filled circles) and field cooled (FC, open circles) measurement of each sample presented in Fig. 1. Close correspondence between ZFC and FC measurements shows negligible difference in the magnetic behavior of the samples on cooling procedures.

**Fig. 4 (color online):** FORC's (left panels) along with corresponding contour plots of the FORC distributions (right panels) for the $x = 0.15$ sample at different temperatures. Contours show a feature extending along $H_c$ at $H_b = 0$ Oe that quickly becomes localized at the origin with increasing temperature.

**Fig. 5:** Projections of the FORC distribution $\rho$ for the $x = 0.15$ sample onto (a) $H_b$, and (b) $H_c$. A monotonic decrease in the peak width and reduction of the magnitude is seen in (a), whereas (b) shows the peak in coercivity moving towards zero.

**Fig. 6 (color online):** FORC's (left panels) along with corresponding contour plots of the FORC distributions $\rho$ (right panels) for the $x = 0.30$ sample at different temperatures. Contours show a negatively bias FORC distribution extending along $H_c$, which becomes localized at the origin with increasing temperature.

**Fig. 7**: Projections of the FORC distribution $\rho$ for the $x = 0.30$ sample onto (a) $H_b$, and (b) $H_c$. (a) shows a monotonic decrease in the peak width and a shift in the peak position from – 0.2 kOe to 0, whereas (b) shows the peak in coercivity moving towards zero.

**Fig. 8:** Fractional irreversible switching obtained from integration of the FORC distribution for the (a) $x = 0.15$ and (b) $x = 0.30$ samples versus temperature. Superimposed is the FM-phase fraction determined from La NMR,[28] showing good agreements between the two.

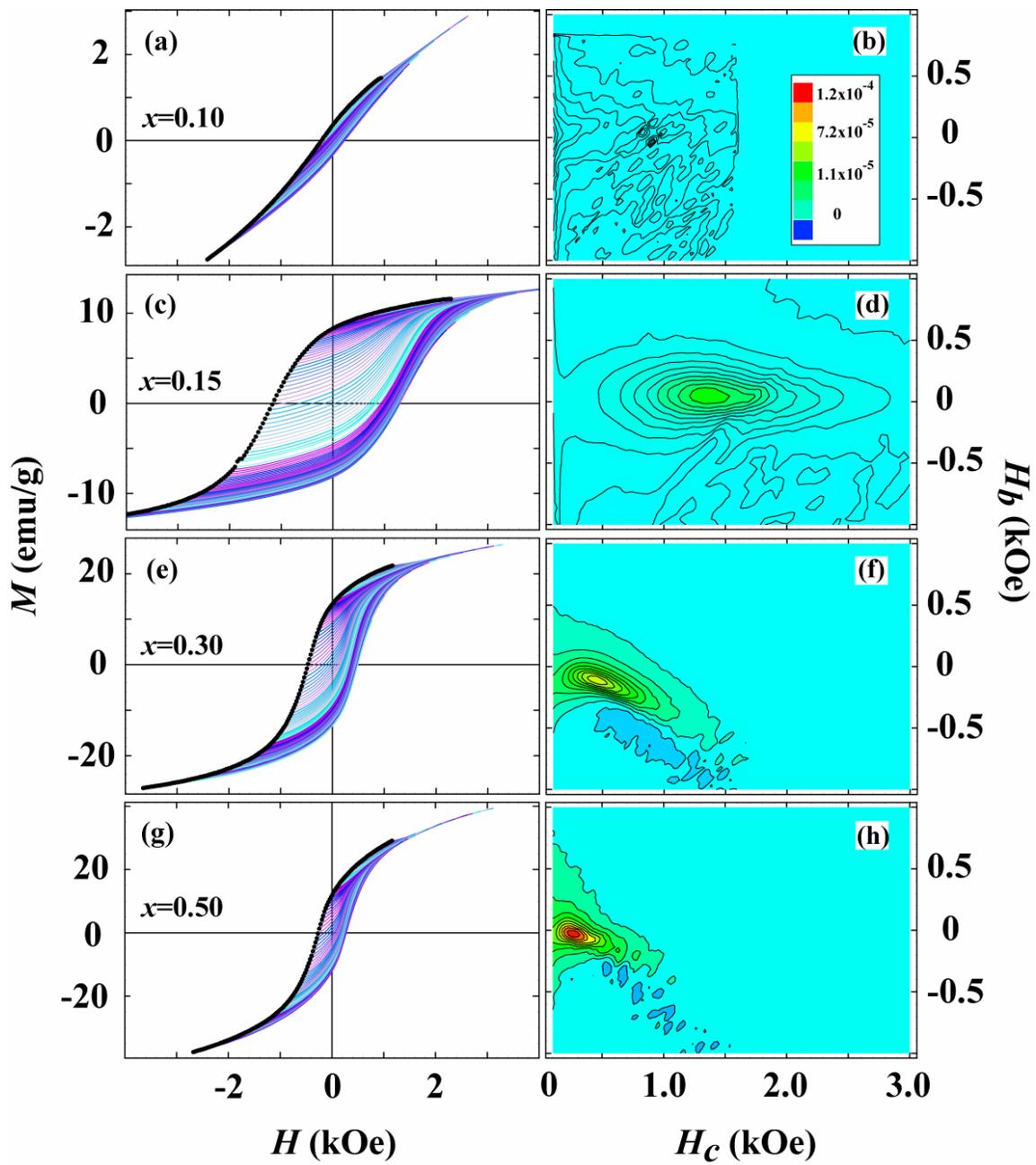

Fig. 1

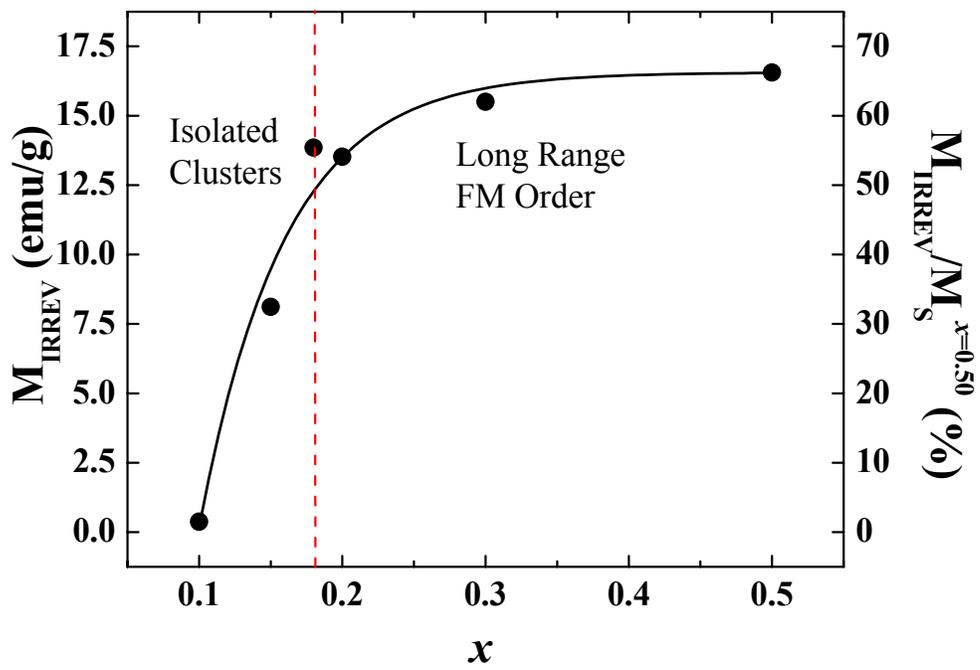

**Fig. 2**

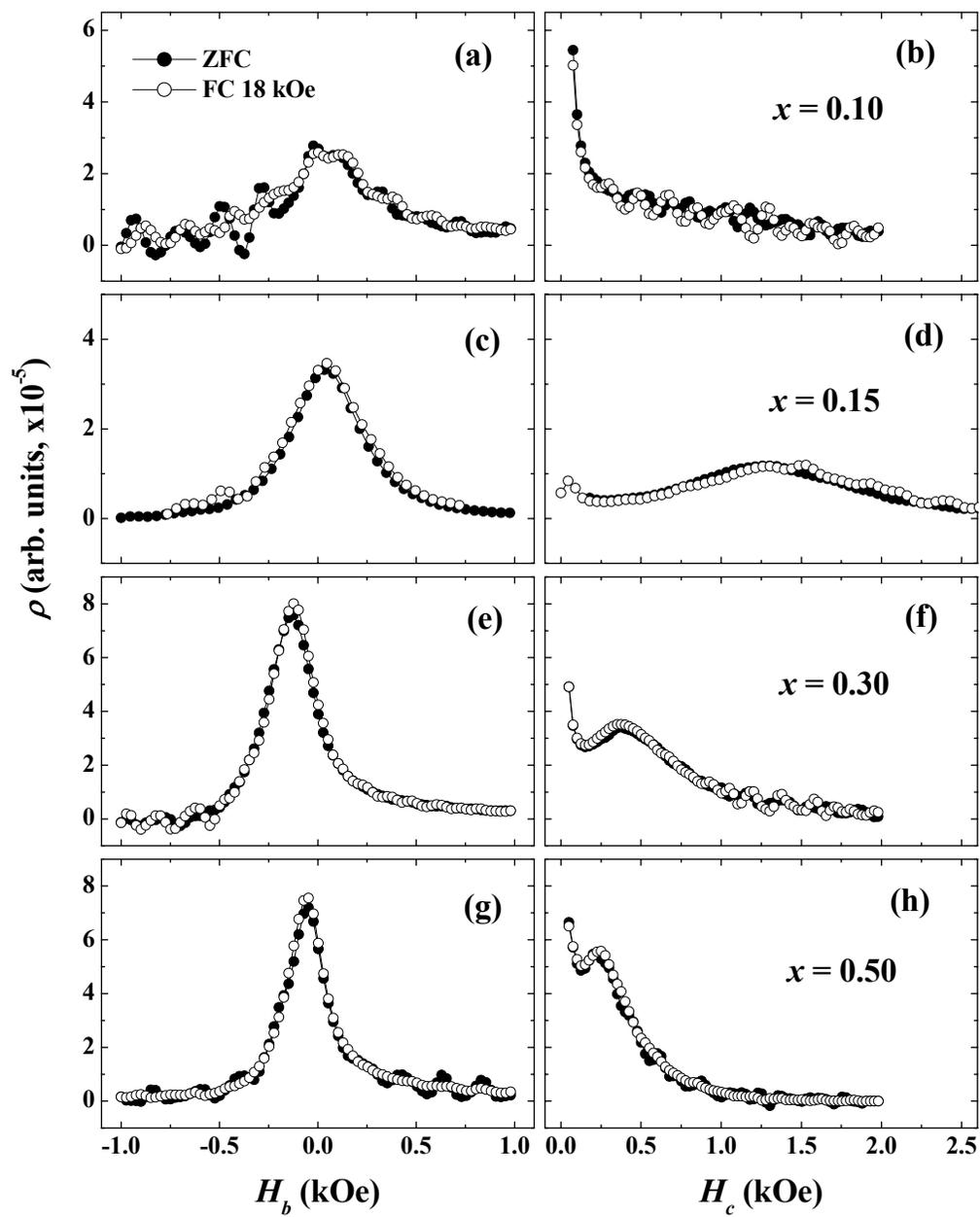

**Fig. 3**

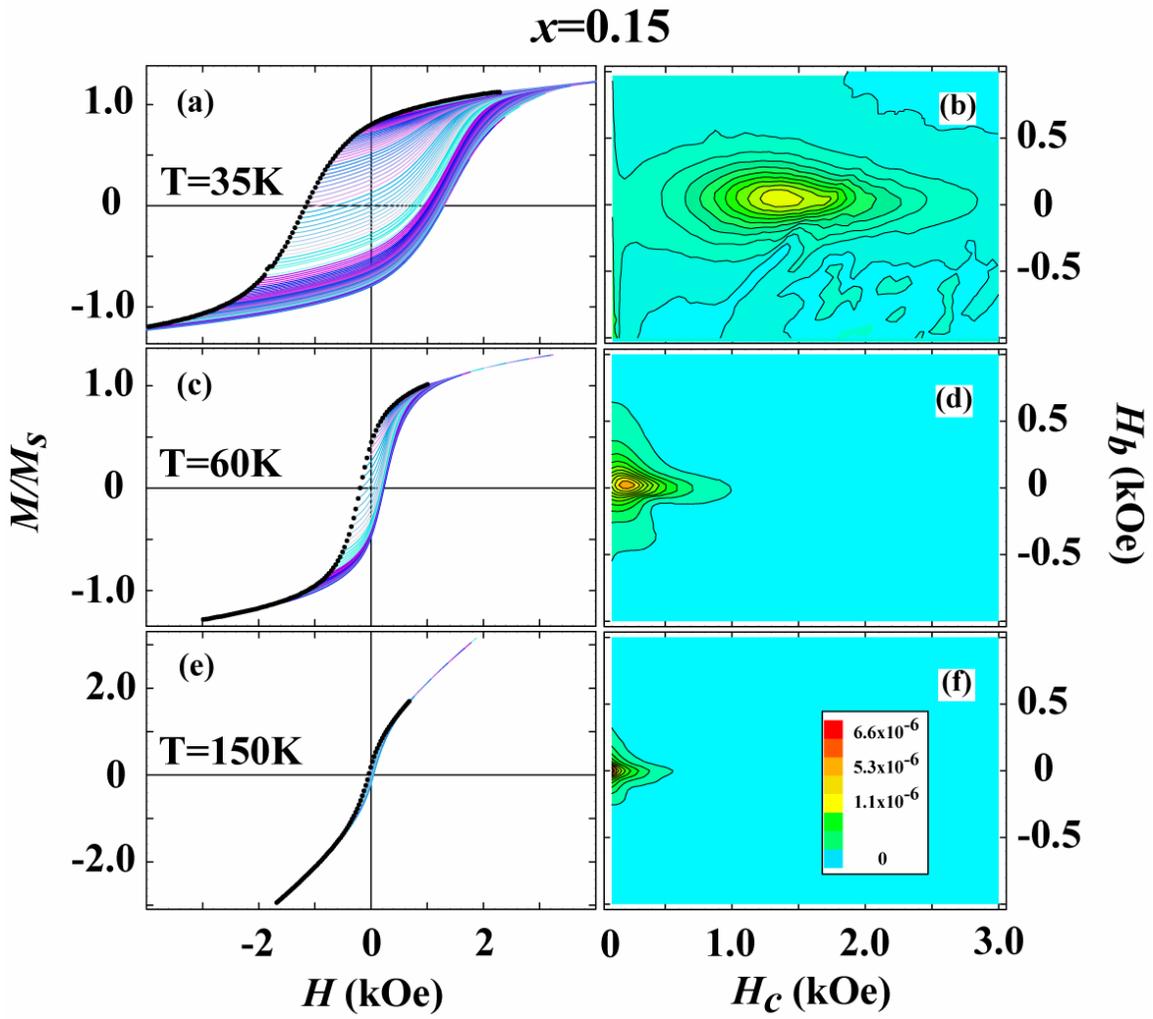

**Fig. 4**

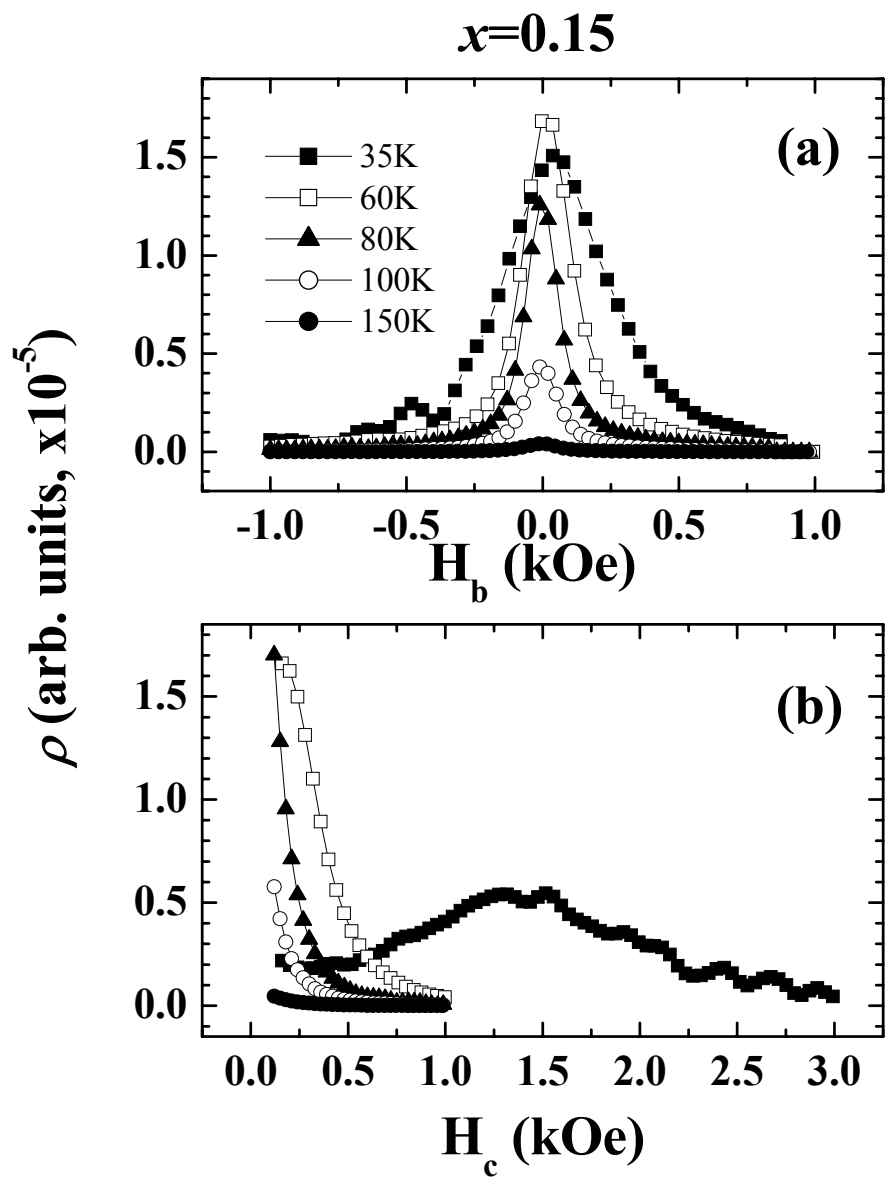

Fig. 5

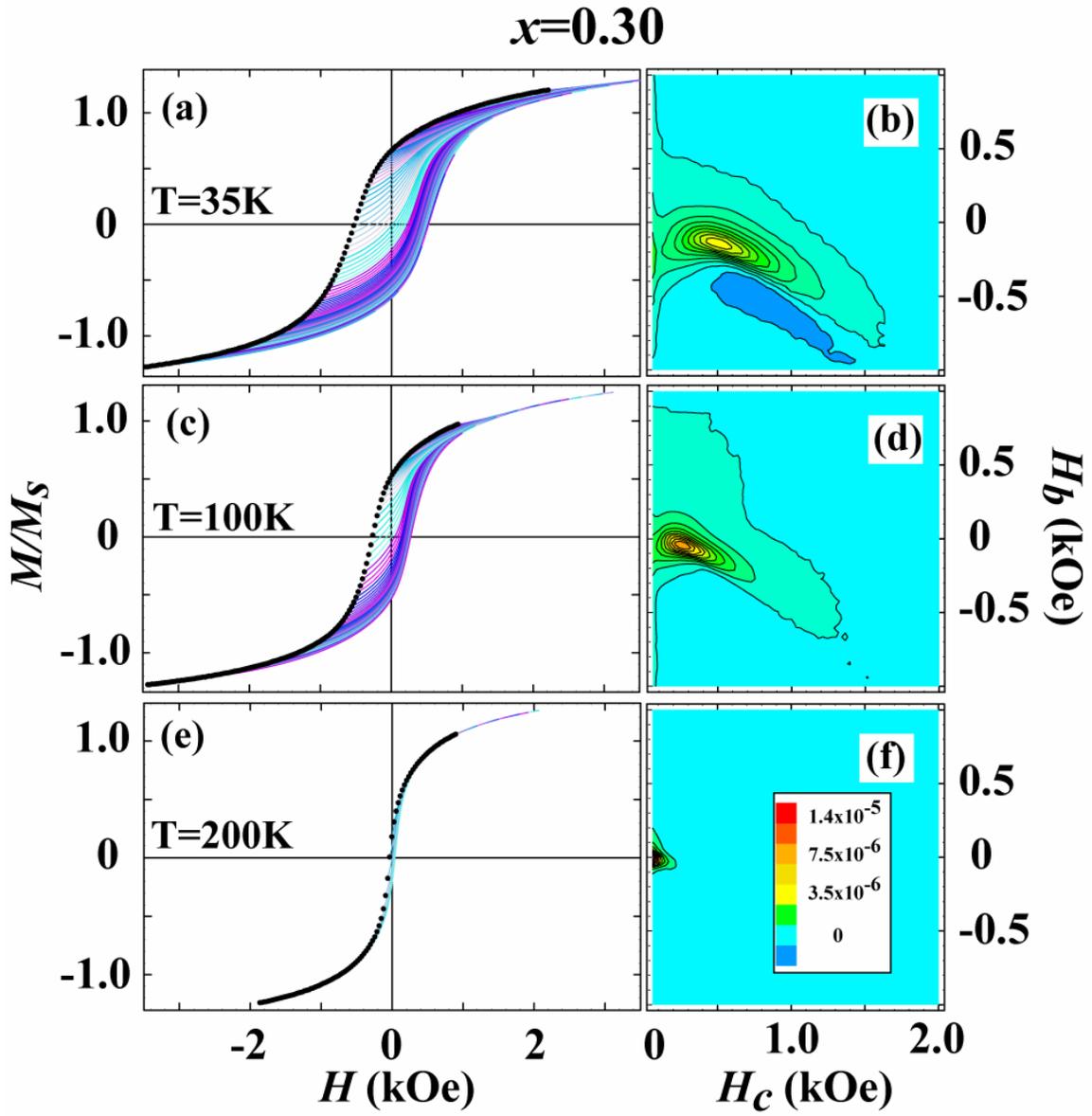

Fig. 6

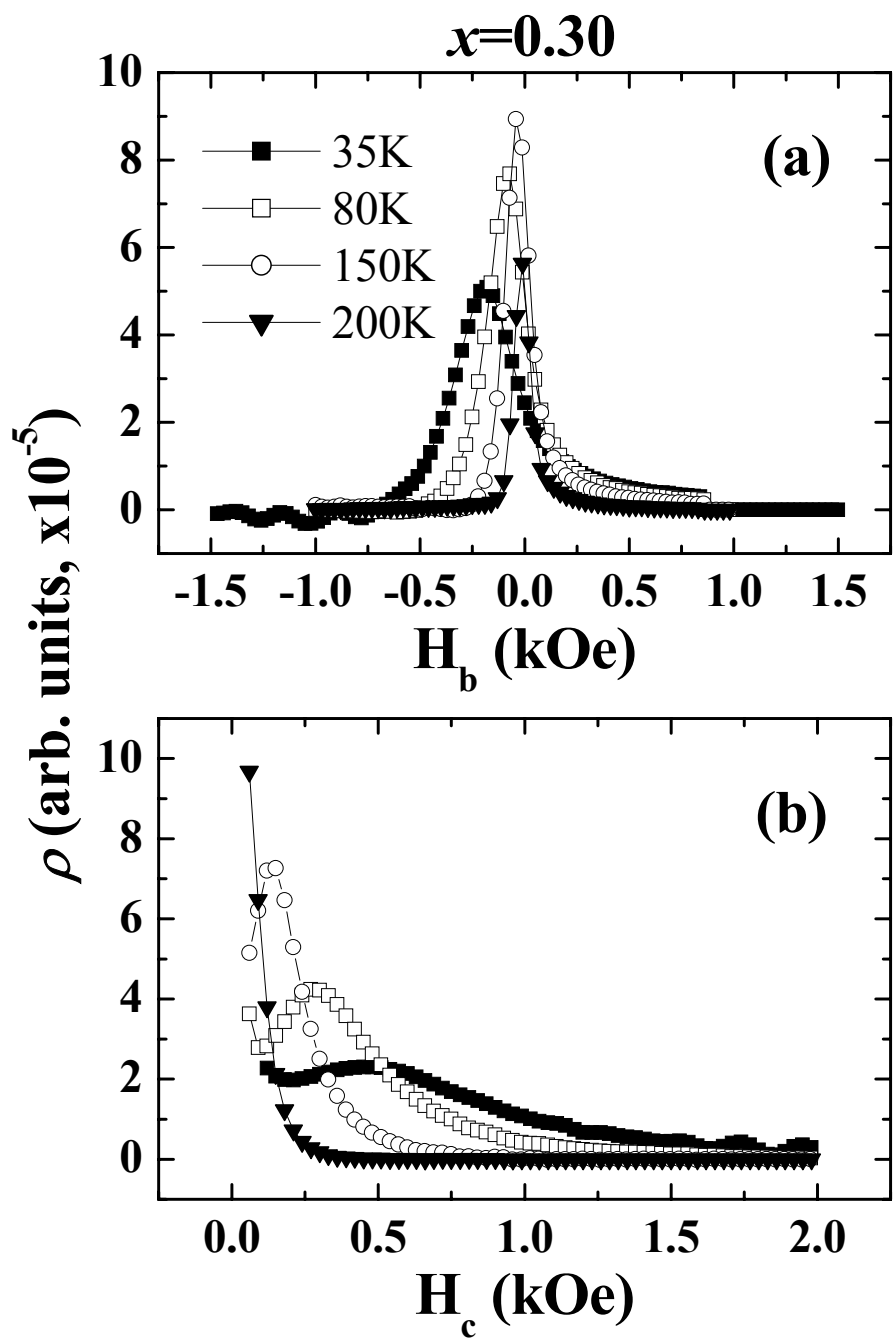

Fig. 7

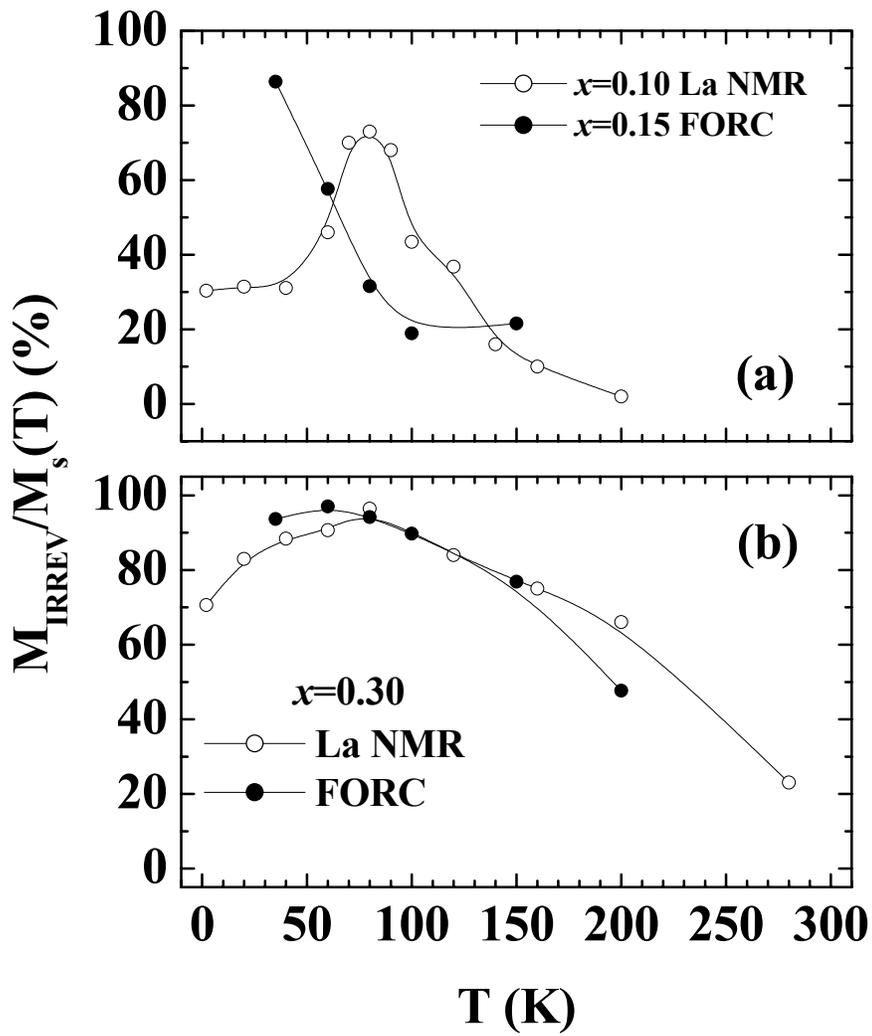

Fig. 8